\documentclass[a4paper]{article}

\newcommand{\be}{\begin{equation}}
\newcommand{\ee}{\end{equation}}
\newcommand{\ba}{\begin{eqnarray}}
\newcommand{\ea}{\end{eqnarray}}
\newcommand{\ban}{\begin{eqnarray*}}
\newcommand{\ean}{\end{eqnarray*}}
\newcommand{\nthm}[2]{\newtheorem}

\newcommand{\braket}[2]{\mbox{$ \langle #1 | #2 \rangle $}}
\newcommand{\sandwich}[3]{\mbox{$ \langle #1 | #2 | #3 \rangle $}}
\newcommand{\ket}[1]{\mbox{$ | #1 \rangle $}}
\newcommand{\bra}[1]{\mbox{$ \langle #1 | $}}

\newcommand{\demi}{\frac{1}{2}}

\newcommand{\compl}{\begin{picture}(8,8)\put(0,0){C}\put(3,0.3){\line(0,1){7}}\end{picture}}

\newcommand{\nat}{\begin{picture}(8,8)\put(0,0){N}\put(0,0){\line(0,1){7}}\end{picture}}
\newcommand{\one}{\begin{picture}(8,8)\put(0,0){1}\put(4.8,0){\line(0,1){7}}\end{picture}}

\newcommand{\ou}{\omega_{1}}
\newcommand{\od}{\omega_{2}}
\newcommand{\oc}{\omega_{c}}
\newcommand{\oz}{\omega_{0}}
\newcommand{\si}{\sigma}

\begin{document}

\title{\Large\sc Quantum Computing}

{\normalsize{\author{
{\bf Valerio Scarani}\\
Institut de Physique Exp\'erimentale\\
Ecole Polytechnique F\'ed\'erale de Lausanne\\
CH-1015 Lausanne, Switzerland\\
valerio.scarani@epfl.ch
}}}
\date{ }
\maketitle

\vspace{2mm}
{\em Am. J. Phys.}, {\bf{66}} (11), November 1998, 
pp. 956-960

\vspace{5mm}

{\footnotesize{Note: this is a revised version September 3rd 1998. The revision is 
minor: I simply modified a misprint (indicated in a footnote) and 
added some references that do not appear in the published versions, 
simply because I have found them after submitting the article. Of 
course, should I write this article today, I would modify some 
paragraphs, and be more sharp in some definitions. But no article 
is perfect, and I think it better to keep the approuved version. I 
gave a talk on QC in Grenoble during the S\'eminaire Daniel 
Dautreppe, September 1998, which is a more systematic review of the field. 
A copy of the PowerPoint file is available by sending me an e-mail.}}

\vspace{5mm}

\begin{abstract}
The main features of quantum computing are described in the framework 
of spin resonance methods. Stress is put on the fact 
that quantum computing is in itself nothing but a re-interpretation 
(fruitful indeed) of well-known concepts. The role of the two basic 
operations, one-spin rotation and controlled-NOT gates, is analyzed, 
and some exercises are proposed.
\end{abstract}

\section{Introduction}

{\em Quantum computing} (QC) is one of the latest booms in science. The 
first detailed paper on QC was published by Deutsch in 1985 \cite{Deu}, but it is
 only in 1994 that Shor 
showed that ``it should work'' \cite{Sho}. Since that date, scientific reviews 
have been filled (and continue to be) with articles related to this 
topic; and an almost entirely new area of theoretical physics has 
been born: the theory of ``quantum error correcting codes''(for a 
simple protocol, see \cite{Div}). Presently 
we still don't know if a quantum computer can be built; but, whatever 
the end of the story may be, I believe that it is worth while working 
out a simple model of quantum computation, and letting 
undergraduate students put it to work on their paper.

\subsection{QC: a new reading of an old book}

One of the striking features of the idea of quantum computer is the 
fact that it contains nothing really new: it is ``nothing but'' a 
re-interpretation of very well-known mathematical objects, mainly the 
theory of quantum two-levels systems. In what follows, we 
shall focus on spins $\frac{1}{2}$, although all that follows could be 
carried out for any two-levels system. Other examples of two-level 
systems which may be of interest can be 
found in \cite{Ben,Wei}. Here you have the translational recipe:
\begin{enumerate}
\item First of all, rename the eigenstates of your 
two-level system as ``0'' and ``1'' (instead of, f.i., 
``spin up'' and ``spin down''): your ``two-levels system'' has become a ``qubit'' 
(the standard shortcut for ``quantum binary digit'').
\item Of course, you must act on your system; don't call it a 
``perturbation'', but a ``logic gate''.
\end{enumerate}
If you have done this, your spin device has 
been transformed in a true one-qubit quantum computer!
Now, a one-bit 
computer is nothing exciting. To obtain a N-qubit computer, you 
have to take N spins, and to be able to adress them in a selective 
way (i.e., you must be able to turn spin number 5, then to couple 
spins number 3,5 and 9, and so on). Nowadays physicists are able to adress 
single quantum states and to work on them; but on systems whose 
{\em decoherence times} are very short. Decoherence is a mechanism we 
are beginning to understand, since some experimental results have 
been obtained \cite{Har}; in a very first approach, decoherence is the 
modification of the quantum state of the system due to interaction 
with an environment. In other words, an irreversible loss of 
information takes place because the system is not perfectly 
isolated \cite{note0}. In the case of a QC, such a mechanism 
can cause the value of a qubit, or the correlation between qubits, to 
change during the calculation, in a way we cannot control. Decoherence 
is the most fundamental obstacle to date preventing us from 
building a QC.
 
Experimental realizations being forbidden up to now, what about calculations? It 
would seem that calculating a quantum computer is a task for a 
computer (a classical one), since one should understand, in 
principle, how a N-qubit logic gate works. However, Barenco {\em et 
al.} \cite{Bar} have shown that 
any possible N-qubit quantum computer operations can be described in 
terms of two basic operations: the rotation of {\em one} spin, and an 
operation involving {\em two} spins, called controlled-NOT (CNOT) or 
exclusive-OR (XOR) gate. This means that from 
the theorist's viewpoint, once you have understood these two simple 
operations, you know everything on how a quantum computer works. 
And, of course, that is exactly what we are going to do in the 
following section.

\subsection{How to rotate one spin}

In this paragraph, we give some basic elements of spin rotation, 
inspired by pulsed Nuclear Magnetic Resonance (NMR) techniques \cite{note1}.

First of all, we define the matrix representation we are going to work 
with, by defining the Pauli matrices as follows:
\be
\begin{array}{ccc}
\sigma_{x}\doteq\left(\begin{array}{cc}0&1\\1&0\end{array}\right),&
\sigma_{y}\doteq\left(\begin{array}{cc}0&-i\\i&0\end{array}\right),&
\sigma_{z}\doteq\left(\begin{array}{cc}1&0\\0&-1\end{array}\right);
\end{array}
\ee
the eigenvectors of $\si_{z}$ are written as 
$\ket{+}_{z}\stackrel{QC}{=}\ket{0}
\doteq\left(\begin{array}{c}1\\0\end{array}\right)$ and 
$\ket{-}_{z}\stackrel{QC}{=}\ket{1}
\doteq\left(\begin{array}{c}0\\1\end{array}\right)$. I find it 
better to keep the NMR notation $\ket{+}_{z}$ until the end of section 
2, to avoid introducing curious terms like ``rotating a qubit around 
an axis''. The reader will be invited to translate these notations 
into QC notations at the beginning of section 3.

It is a matter of evidence that any rotation can be decomposed using 
only rotations around $\hat{e}_{z}$ and (say) $\hat{e}_{x}$ (a more 
precise statement is given in \cite{Bar}, Lemma 4.1). So we 
will simply give one-spin Hamiltonians allowing to perform these two 
rotations. Here we remind the general form of rotation matrices in our 
representation, with the convention that clockwise rotation is positive 
(we give also $R_{y}(\theta)$, which may be useful in 
practice)
\ba
R_{x}(\theta)\doteq\left(\begin{array}{cc}\cos\theta&i\sin\theta\\
i\sin\theta&\cos\theta\end{array}\right),&
R_{y}(\theta)\doteq\left(\begin{array}{cc}\cos\theta&\sin\theta\\
-\sin\theta&\cos\theta\end{array}\right),&
R_{z}(\theta)\doteq\left(\begin{array}{cc}e^{i\theta}&0\\
0&e^{-i\theta}\end{array}\right).
\ea  

We consider now a spin $\demi$ in a static uniform magnetic field 
$\vec{B_{0}}=B_{0}\hat{e}_{z}$; we suppose that this spin feels another 
(much weaker) interaction, such that the total Hamiltonian at 
equilibrium is
\be
H_{0}=-\frac{\hbar}{2}(\omega_{0}+\omega_{1})\si_{z}
\ee
with $\omega_{0}=\gamma B_{0}$ the Larmor frequency associated to the 
external field. The external field plays here a most trivial role: 
basically, we need it to lift by a sufficient amount the degeneracy 
of spin levels; so it is customary to work in a rotating frame in 
which the contribution of $\vec{B}_{0}$ cancels out. This frame is 
defined by:
\ba
\hat{e}_{x'}=\cos\omega_{0} t\,\hat{e}_{x}+\sin \omega_{0} t\,\hat{e}_{y}\\
\hat{e}_{y'}=-\sin \omega_{0} t\,\hat{e}_{x}+\cos \omega_{0} t\,\hat{e}_{y}\\
\hat{e}_{z'}=\hat{e}_{z}
\ea
and the static Hamiltonian becomes simply
\be
H_{0}'=-\frac{\hbar}{2}\omega_{1}\si_{z}.
\ee
Under $H_{0}'$, the most general spin state 
$\alpha\ket{+}_{z}+\beta\ket{-}_{z}$ evolves according to
\be
\ket{\psi}(t)\doteq\left(\begin{array}{c}\alpha(t)\\\beta(t)\end{array}\right)=
\left(\begin{array}{cc}e^{i\frac{\ou t}{2}}&0\\
0&e^{-i\frac{\ou t}{2}}\end{array}\right)
\left(\begin{array}{c}\alpha(0)\\\beta(0)\end{array}\right)\doteq
R_{z}(\frac{\ou t}{2})\ket{\psi}(0)
\ee
so the ``free'' evolution gives us the possibility of performing 
rotations around $\hat{e}_{z}$. Rotations around $\hat{e}_{x'}$ 
can be obtained by applying the time-dependent Hamiltonian
\be
H_{pert}=-\frac{\hbar}{2}\omega_{p}
\Big[\cos(\oz+\ou)t\si_{x}+\sin(\oz+\ou)t\si_{y}\Big]
\ee
which yields for any practical purpose \cite{note3}
\be
\ket{\psi}(t)\doteq
\left(\begin{array}{c}\alpha(t)\\\beta(t)\end{array}\right)=
\left(\begin{array}{cc}\cos\frac{\omega_{p}t}{2}&i\sin\frac{\omega_{p}t}{2}\\
i\sin\frac{\omega_{p}t}{2}&\cos\frac{\omega_{p}t}{2}\end{array}\right)
\left(\begin{array}{c}\alpha(0)\\\beta(0)\end{array}\right)
\doteq R_{x'}(\frac{\omega_{p}t}{2})\ket{\psi}(0).
\label{eq:evol}
\ee

Before turning to the model for QC, it is important 
to notice that:
\begin{enumerate}
\item The energy levels separation at equilibrium plays an important 
role; by adjusting the {\em frequency} of the perturbation (above, 
$\oz+\ou$), one can select one transition in a multilevel system.
\item A rotation of the state \cite{note1b} 
by an angle $\theta$ around $\hat{e}_{x'}$ is 
obtained by applying the perturbation during a time 
$\tau_{\theta}=\frac{2\theta}{\omega_{p}}$, depending on the {\em 
intensity} of the perturbation. 
\end{enumerate}
We shall make extensive use of the first remark in what follows. Here 
we must tell something more on the second remark. We have said that 
spin rotation is obtained by applying during a well-defined time $\tau$ a pulse having a 
well-defined frequency $\omega_{r}$. Such a pulse does not excite only 
resonance at $\omega_{r}$; it excites a frequency band 
$\omega_{r}\pm\Delta\omega$, where basically $\Delta\omega\sim\tau^{-1}$ 
\cite{note2}. This means that by lengthening\footnote{In the published 
version I wrote by inadvertance ``shortening'', which is of course wrong.} 
the pulse 
(i.e., by increasing the intensity of the perturbation), one can be 
more selective; and viceversa.

\section{Putting a Quantum Computer to work!}

\subsection{The model}

The system we work with are two spins $\frac{1}{2}$ (two qubits); the 
Hilbert space describing such a system is 
${\cal{H}}=\compl^2\otimes\compl^2$, the tensor product of two copies 
of $\compl^2$, each describing one spin \cite{note4}. The static Hamiltonian 
will be taken as
\be
H_{0}=-\frac{\hbar}{2}\Big[\Omega_{1}\,(\sigma_{z}^{1}\otimes\one)
+\Omega_{2}\,(\one\otimes\sigma_{z}^{2})
+\omega_{c}\,(\sigma_{z}^{1}\otimes\sigma_{z}^{2})\Big]
\label{eq:hamilt}
\ee
whose eigenstates are the four products of two Pauli matrices 
eigenstates. In all that follows we shall use notations like 
$\ket{++}$ as shortcuts for $\ket{+}_{z}\otimes\ket{+}_{z}$. We used 
the notation $\Omega_{i}=\oz+\omega_{i}$. We choose the following 
representation: 
$\ket{++}\doteq e_{1}$, $\ket{-+}\doteq e_{2}$, $\ket{+-}\doteq e_{3}$, 
$\ket{--}\doteq e_{4}$; where of course $e_{i}$ is the 
column four-tuple whose elements are: $1$ at the $i$th place and $0$ at 
the others. We have then
\be
H_{0}\doteq
-\frac{\hbar}{2}\left(\begin{array}{cccc}\Omega_{1}+\Omega_{2}+\omega_{c}&&&\\
&-\omega_{1}+\omega_{2}-\omega_{c}&&\\&&\omega_{1}-\omega_{2}-\omega_{c}&\\
&&&-\Omega_{1}-\Omega_{2}+\omega_{c}
\end{array}\right).
\ee
We choose $\omega_{1}>\omega_{2}$, and we assume that 
$\omega_{c}<<\oz$ (weak coupling) and that $\ou-\od\geq 4\oc$ (the 
reason for this is given in a subsequent discussion). 
The energy levels diagram is immediately drawn, whence we can easily 
derive the transition frequency spectrum, drawn in Fig.1 (the low frequency 
transition $\ket{+-}\leftrightarrow\ket{-+}$ and the high 
frequency transition $\ket{++}\leftrightarrow\ket{--}$ are omitted, since 
we are not interested in transitions involving both spins). 
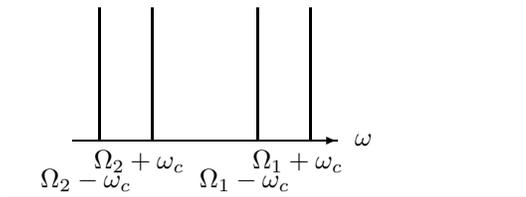
\begin{center}
\begin{picture}(150,90)
\thicklines
\put(10,15){\line(0,1){50}}
\put(30,15){\line(0,1){50}}
\put(70,15){\line(0,1){50}}
\put(90,15){\line(0,1){50}}
\put(110,15){\makebox(0,0){$\omega$}}
\put(5,0){\makebox(0,0){$\Omega_{2}-\oc$}}
\put(25,7){\makebox(0,0){$\Omega_{2}+\oc$}}
\put(65,0){\makebox(0,0){$\Omega_{1}-\oc$}}
\put(85,7){\makebox(0,0){$\Omega_{1}+\oc$}}
\thinlines
\put(0,15){\vector(1,0){100}}
\end{picture}

{\em Figure 1:} Transition frequency spectrum for hamiltonian (\ref{eq:hamilt}).
\end{center}

Let's discuss some conditions:
\begin{enumerate}
\item To perform {\em one-spin rotation} around an axis 
lying in the $(\hat{e}_{x},\hat{e}_{y})$ plane (remember that 
rotations around $\hat{e}_{z}$ are obtained by letting the system evolve 
under the static Hamiltonian, and do not involve any resonance 
technique) on spin 1, we 
must be able to adress {\em both} $\ket{++}\leftrightarrow\ket{-+}$ and 
$\ket{+-}\leftrightarrow\ket{--}$ without excitating any other 
transition; an analogue requirement must be satisfied for one-spin 
rotation on spin 2. This yields 
a condition on the physical parameters, namely $\ou-\od>2\oc$ (to work 
more confortably, when we anticipated this condition we took $4\oc$ as upper 
bound) and an upper limit for $\tau_{\theta}$ at fixed $\theta$ (for 
this operation, the pulse must not be too selective).
\item We shall see that to perform all the possible {\em CNOT operations} 
means the possibility of adressing each transition separately; this 
yields a lower bound for $\tau_{\theta}$ at fixed $\theta$ (for this 
operation, we need selective pulses).
\end{enumerate}
In all that follows, we suppose that we are able to control the 
frequency and intensity of each pulse, in order to adress the chosen 
transition with the desired selectivity.

\subsection{Rotations and CNOT (XOR) gates}

The discussion of {\em one-spin rotations} is merely a matter of 
re-writing, since we know everything thereabout. Writing
\be
R_{u}(\theta)\doteq
\left(\begin{array}{cc}r_{11}&r_{12}\\r_{21}&r_{22}\end{array}\right)
\ee
one obtains immediately for our representation on 
$\compl^2\otimes\compl^2$:
\ba
R_{u}(\theta)\otimes\one\doteq\left(\begin{array}{cc}R_{u}(\theta)&0\\ 0&R_{u}(\theta)
\end{array}\right);\\
\one\otimes R_{u}(\theta)\doteq\left(\begin{array}{cc} 
r_{11}\one&r_{12}\one\\ r_{21}\one&r_{22}\one
\end{array}\right).
\ea
Thus we must now focus our attention on {\em CNOT gates}. We recall 
that this stands 
for {\em controlled NOT}, and means that we flip one spin according 
to the state of the other. The following matrix is a CNOT, in which if 
spin 2 is in the state $\ket{-}$ then the state of spin 1 is flipped:
\be
C_{2-}^{1}\doteq
\left(\begin{array}{cccc}1&0&&\\0&1&&\\&&0&1\\&&1&0\end{array}\right).
\label{eq:cnot}
\ee
It is not hard to describe this operation using our tools: all we need 
to do, is to adress uniquely the $\ket{+-}\leftrightarrow\ket{--}$ 
transition, with a pulse whose lenght $\tau$ is defined by [see 
(\ref{eq:evol})] $\frac{\omega_{p}\tau}{2}=\frac{\pi}{2}$. The reader 
is invited to write down the other three possible CNOT gates.

Note that one-spin rotations are intrinsically non-classical, since 
(in general) they generate superposition states. On the 
contrary, the CNOT operation is in itself classical; however, in a QC we want 
to perform such an operation on arbitrary states, and this assumes 
highly non-classical features. So for instance:
\ba
C_{2-}^{1}\big(\frac{1}{\sqrt{2}}\ket{++}+\frac{1}{\sqrt{2}}\ket{--}\big)&=
\frac{1}{\sqrt{2}}\ket{++}+\frac{1}{\sqrt{2}}\ket{+-}=\nonumber\\
&=\ket{+}\otimes\big(\frac{1}{\sqrt{2}}\ket{+}+\frac{1}{\sqrt{2}}\ket{-}\big)
\ea
which means {\em disentanglement}. Some short exercises follow, whose 
purpose is twofold: understanding the different role of one-spin 
rotations and CNOT gates in a quantum computation; and ``feeling'' (a 
general proof is not our purpose here) that with these two operations 
one can simulate any quantum calculation.

\section{Exercises}

Even though this is totally trivial, the reader is invited to 
``translate'' spin states into ``qubits'' using the standard rules of 
binary calculations; thus f.i.
\ban
\ket{++}=\ket{00}=\ket{0},&
\ket{-+}=\ket{10}=\ket{1}\\
\ket{+-}=\ket{01}=\ket{2},&
\ket{--}=\ket{11}=\ket{3}.
\ean
This translation is used in Exercises 3 and 4.

\subsection{Three-spins maximally entangled state (GHZ)}
Give an algorithm using only one-spin rotations and CNOT gates to 
transform the fundamental three-spins state $\ket{+++}$ into the 
maximally entangled GHZ state $\frac{1}{\sqrt{2}}\big(\ket{+++}+\ket{---}\big)$.
 
Imagine now you {\em don't know} the input state: do you have any 
hope of building a ``universal GHZ preparator'', i.e. an algorithm 
that transforms {\em any input state whatsoever} into the GHZ state?

{\em Solution}

Here is a possible sequence starting from $\ket{+++}$:
\ban
\Big[\one\otimes\one\otimes R_{y}(\frac{\pi}{4})\Big]\ket{+++}=
\frac{1}{\sqrt{2}}\big(\ket{+++}+\ket{++-}\big)\\
\Big[\one\otimes C_{3-}^{2}\Big]\frac{1}{\sqrt{2}}\big(\ket{+++}+\ket{++-}\big)=
\frac{1}{\sqrt{2}}\big(\ket{+++}+\ket{+--}\big)\\
\Big[C_{2-}^{1}\otimes\one\Big]\frac{1}{\sqrt{2}}\big(\ket{+++}+\ket{+--}\big)=
\frac{1}{\sqrt{2}}\big(\ket{+++}+\ket{---}\big)
\ean
Of course, it is {\em not} possible to find an algorithm that gives the 
same output state (GHZ, or whatever else) for any input state: 
QC is concerned with unitary evolution, thus in particular orthogonal 
input states must give orthogonal output states.

\subsection{The NOT logic gate}

Write down the matrix representation of the NOT logic gate (inversion 
of all spins) for a two-spins system. What are its eigenstates? Can 
such a gate modify entanglements?

{\em Solution}

The NOT logic gate is
\be
N\doteq\left(\begin{array}{cccc}&&0&1\\&&1&0\\0&1&&\\1&0&&\end{array}\right).
\ee
One can write it down either by direct reasoning on the four basis 
states, or by calculating the product of two one-spin NOTs. Since 
$N=-\big(R_{x}(\frac{\pi}{2})\otimes\one\big)
\big(\one\otimes R_{x}(\frac{\pi}{2})\big)$ is (up to an overall 
phase factor) the product of two one-spin rotations, such a gate 
cannot modify entanglements. Its eigenstates form the so-called Bell 
basis:
\ba
\ket{\Phi^{\pm}}=\frac{1}{\sqrt{2}}\big(\ket{++}\pm\ket{--}\big),&
\ket{\Psi^{\pm}}=\frac{1}{\sqrt{2}}\big(\ket{+-}\pm\ket{-+}\big).
\label{eq:bell}
\ea

\subsection{Readout of Bell states}

This exercise is inspired 
by \cite{Zei}: had Zeilinger's group had a suitable logic gate for 
their polarized photons, the readout would have been by far easier!
Imagine thus you have done an experiment whose result is one of the four 
Bell states (\ref{eq:bell}). However, your detectors' eigenstates 
are {\em not} Bell states, but the standard basis states $\ket{++}$ 
etc. Write down a logic gate which would permit you to make the 
translation, in matrix representation and as a product of basic 
operations (suggestion: first destroy entanglement, then 
superpositions). 

{\em Solution}

We choose the following translation:
\ban
\ket{\Phi^{+}}\Longrightarrow\ket{0},&\ket{\Psi^{+}}\Longrightarrow\ket{1}\\
-\ket{\Phi^{-}}\Longrightarrow\ket{2},&-\ket{\Psi^{-}}\Longrightarrow\ket{3}.
\ean
Thus the logic gate will be
\be
T\doteq\frac{1}{\sqrt{2}}
\left(\begin{array}{cccc}1&0&0&1\\0&1&1&0\\-1&0&0&1\\0&-1&1&0\end{array}\right).
\ee
It is not difficult to decompose this gate into a product of the basic 
operations. One possible solution is $T=\big(\one\otimes 
R_{y}(\frac{\pi}{4})\big)C_{2-}^{1}$.

\subsection{Quantum Fourier Transform}
The operator known as {\em quantum Fourier transform} plays an 
important role in Shor's algorithm \cite{Sho,Ben}. 
For a system of $n$ spins\ldots sorry, $n$ qubits, it is defined 
as (we write $Q=2^n$)
\be
F=\frac{1}{\sqrt{Q}}\sum_{x,k=0}^{Q-1}\ket{k}e^{2\pi ikx/Q}\bra{x}.
\ee
Verify that $F$ is unitary. Write down the matrix representation of 
$F$ for $n=2$.

{\em Solution}

One has
\ban
F^\dag F&=&\frac{1}{Q}\sum_{x,k,x',k'}
\ket{x'}e^{-2\pi ik'x'/Q}\underbrace{\braket{k'}{k}}_{\delta_{k,k'}}e^{2\pi ikx/Q}\bra{x}=\\
&=&\sum_{x,x'}\ket{x'}
\underbrace{\Big(\frac{1}{Q}\sum_{k=0}^{Q-1}\exp\big(2\pi 
ik(x-x')/Q\big)\Big)}
_{\sandwich{x'}{F^\dag F}{x}}\bra{x}.
\ean
If $x=x'$, then $\sandwich{x'}{F^\dag F}{x}=1$. Otherwise,
\[
\sandwich{x'}{F^\dag F}{x}=\frac{1}{Q}
\frac{1-e^{2\pi i(x-x')}}{1-e^{2\pi i(x-x')/Q}}=0
\]
since $\nat\ni|x-x'|<Q$. Thus $F^\dag F=\one$.

For $n=2$ one has
\be
F\doteq\demi\left(\begin{array}{cccc}1&1&1&1\\1&i&-1&-i\\
1&-1&1&-1\\1&-i&-1&i\end{array}\right).
\ee

\section{Conclusion}
In this paper, the reader has found a self-contained description of a 
quantum computer based on well-known elements of undergraduate 
quantum mechanics. I wanted to stress that, while the connected field 
of quantum error correcting codes is something for specialists (and that's 
why it was totally neglected here), the basic idea of quantum computation 
is something very simple; so simple that most readers have probably 
already made many ``quantum calculations'' without calling them by 
this name!

Some more references that do not appear in the published version:
\begin{itemize}
\item Laflamme, Knill, Zurek, Catasti, Mariappan, quant-ph/9709025, to 
appear in Proc. Roy. Soc. Lond. (NMR realization of a GHZ state)
\item Chuang, Gershenfeld, Kubinec, {\em Phys. Rev. Lett.} {\bf{80}} (1998) 
3408; and Jones, Mosca, Hansen, quant-ph/9805069 (NMR experimental 
realization of Grover's algorithm)
\item Chuang, Vandersypen, Zhou, Leung, Lloyd, {\em Nature} 
{\bf{393}} (1998) 143-146 (NMR experimental realization of 
Deutsch-Josza algorithm)
\item Haroche, Raimond, {\em Phys. Today} 
August 1996, 51 (about the problems of building a QC ``dream or 
nightmare?'')
\item Steane, quant-ph/9708022 (a review article, with a good section 
about information theory)
\item Cory, Fahmy, Havel, {\em Proc. Natl. Acad. Sci. USA} {\bf{94}} 
(1997) 1634 (almost the same as ref [11], published at the same time 
independently)
\item Cirac, Zoller, {\em Phys. Rev. Lett.} {\bf{74}} (1995) 4091 (the 
basic proposal for trapped ions QC, a promising technique not yet 
realized)
\item Kane, {\em Nature} {\bf{393}} (1998) 133-137 (a proposal of 
implementation with nuclear spins in a solid-state device) 

\end{itemize}


\begin{thebibliography}{9}
\bibitem{note1} Here you have three good reasons for this choice: (1) NMR is a 
standard experimental technique, contrarily f.i. to Cavity QED: many 
physicists and chemists are used to NMR, and those who are not can 
find excellent descriptions in many standard books (see f.i. 
\cite{Coh}). (2) As discussed by Gershenfeld 
and Chuang \cite{Ger}, NMR is up to now the only proposed method for 
QC involving very long decoherence times. (3) NMR is my own research field, 
and I am accostumed to explain it to undergraduate students!
\bibitem{note3} This is rigorously 
true in another frame, rotating with 
angular frequency $\oz+\ou$. But we assume that $\omega_{p}>>\ou$, 
i.e., the induced rotation around $\hat{e}_{x'}$ is much more rapid 
than the ``free'' rotation around $\hat{e}_{z}$. Since we will be 
interested in small angles ($\omega_{p}t=\pi$ at most), we can consider that no 
free evolution takes place while the perturbation is applied.
\bibitem{note1b} Let's remind the reader as an example that the spin flip 
$\ket{+}_{z}\leftrightarrow\ket{-}_{z}$ is obtained by rotating the state vector 
around an axis perpendicular to $\hat{e}_{z}$ by $\frac{\pi}{2}$, 
and not by $\pi$.
\bibitem{note2} The reader can 
check it out as a complementary exercise by calculating the Fourier 
transform of a rectangular pulse: \[ p(t)=\left\{\begin{array}{cc}
\sin\omega_{r} t,&0<t<\tau\\0&\mbox{otherwise}
\end{array}\right. .\]
\end{thebibliography}
\end{document}